\documentclass[11pt]{article}
\usepackage{graphicx}

\newcommand{\BABARPubYear}    {05}

\newcommand{\BABARProcNumber} {130}
\newcommand{\SLACPubNumber} {11594}

\input babarsym

\long\def\inst#1{\par\nobreak\kern 4pt\nobreak
    {\it #1}\par\vskip 10pt plus 3pt minus 3pt}

\RequirePackage{xspace}
\def\C       {\ensuremath{C}\xspace}

\def\CP            {\ensuremath{C\!P}\xspace}
\def\B             {\ensuremath{B}\xspace}
\def\bccs          {\ensuremath{b\rightarrow\c\cbar\s}}
\def\bccd          {\ensuremath{b\rightarrow\c\cbar\d}}

\def\bsqq          {\ensuremath{b\rightarrow\s\qbar\q}}
\def\sintwob       {\ensuremath{\sin 2 \beta}}
\def\costwob       {\ensuremath{\cos 2 \beta}}

\def\btojpsipi     {\ensuremath{\B^0 \to {J}/\psi\:\piz}}

\def\btodstardstar {\ensuremath{\B^0 \to {D^{*+}}{D^{*-}}}}
\def\btojpsikstar  {\ensuremath{\B^0 \to {J}/\psi\:{K}^{\star}}}
\def\btoetaprimekz {\ensuremath{\B^0 \to \etaprime\kz}}
\def\btoetaprimekl {\ensuremath{\B^0 \to \etaprime\kl}}
\def\btoetaprimeks {\ensuremath{\B^0 \to \etaprime\ks}}
\def\btopizkz      {\ensuremath{\B^0 \to \piz\kz}}
\def\btopizpizkz   {\ensuremath{\B^0 \to \piz\piz\kz}}
\def\btofzkz       {\ensuremath{\B^0 \to \fz\kz}}
\def\btophikz      {\ensuremath{\B^0 \to \phi\kz}}
\def\btoomegakz    {\ensuremath{\B^0 \to \omega\kz}}
\def\btokpkmkz     {\ensuremath{\B^0 \to \kpkmkz}}

\def\btoksksks     {\ensuremath{\B^0 \to \ksksks}}

\def\etacks        {\ensuremath{ \etac{K}_{S}}}
\def\chiks         {\ensuremath{ \chione{K}_{S}}}

\def\kstardecay    {\ensuremath{{K}^{\star}\rightarrow\ks\piz}}
\def\jpsikstar     {\ensuremath{{J}/\psi\:{K}^{\star}}}

\def\jpsi          {\ensuremath{J}/\psi}
\def\etac {\ensuremath\eta_{c}}
\def\etaf {\ensuremath\eta_{\f}}
\def\chione {\ensuremath\chi_{c1}}

\def\etaprime {\ensuremath{{\eta}^{'}}}

\def\ksksks {\ensuremath{\ks\ks\ks}}

\def\kpkmkz {\ensuremath{\kp\km\kz}}

\def\kp   {\ensuremath{{K}^+}}
\def\km   {\ensuremath{{K}^-}}
\def\kz   {\ensuremath{{K}^0}}
\def\ks   {\ensuremath{{K}_{S}}}
\def\kl   {\ensuremath{{K}_{L}}}
\def\fz   {\ensuremath{ {\f}_{0}}}
\def\piz  {\ensuremath{ \pi^0 }}
\def\pip  {\ensuremath{ \pi^+ }}
\def\pim  {\ensuremath{ \pi^- }}

\def\CP {\ensuremath{CP}}
\def\C {\ensuremath{C}}
\def\S {\ensuremath{S}}

\def\Bcp {\ensuremath{B}_{\CP}}
\def\Btag {\ensuremath{B_{tag}}}

\def\tcp  {\ensuremath{t_{\CP}}}
\def\ttag {\ensuremath{t_{tag}}}

\def\dm   {\ensuremath{\Delta m}}

\def\dz   {\ensuremath{\Delta z}}

\def\B    {\ensuremath{B}}
\def\f    {\ensuremath{f}}
\def\b    {\ensuremath{b}}
\def\t    {\ensuremath{t}}
\def\d    {\ensuremath{d}}

\def\s    {\ensuremath{s}}
\def\c    {\ensuremath{c}}
\def\q    {\ensuremath{q}}

\def\qbar {\ensuremath{\overline{q}}}
\def\cbar {\ensuremath{\overline{c}}}

\def\bb   {\ensuremath{B \overline{B} }}
\def\Bz {\ensuremath{B^{0}}}
\def\bzero {\ensuremath{B^{0}}}

\def\Bzb {\ensuremath{\overline{B^{0}}}}

\def\Acp {\ensuremath{A_{CP} }}

\usepackage{relsize}
\def\babar{\mbox{\slshape B\kern-0.1em{\smaller A}\kern-0.1em
    B\kern-0.1em{\smaller A\kern-0.2em R}}}

\begin{document}
{\pagestyle{empty}

\begin{flushright}
SLAC-PUB-\SLACPubNumber \\
\babar-PROC-\BABARPubYear/\BABARProcNumber \\
December 2005\\
\end{flushright}

\par\vskip 4cm

\begin{center}
\Large \bf Measurements of $\sintwob$ at \babar\ with charmonium and penguin decays.
\end{center}
\bigskip

\begin{center}
\large 
Katherine George\\
University of Liverpool\\
Department of Physics, Oxford Street, Liverpool L69 7ZE, United~Kingdom.\\
(for the \lbabar\ Collaboration)
\end{center}
\bigskip \bigskip

\begin{center}
\large \bf Abstract
\end{center}
This article summarises measurements of time-dependent $\CP$ asymmetries in
decays of neutral \B\ mesons to charmonium, open-charm and gluonic penguin-dominated
charmless final states. Unless otherwise stated, these measurements are based on a
sample of approximately 230 million $\Upsilon$(4S)$\rightarrow\bb$ decays collected by
the \babar\ detector at the PEP-II asymmetric-energy \B-factory.

\vfill
\begin{center}
Contributed to the Proceedings of PANIC05 - Particles and Nuclei International Conference, \\
Santa Fe, NM - October 24-28, 2005 
\end{center}

\vspace{1.0cm}
\begin{center}
{\em Stanford Linear Accelerator Center, Stanford University, 
Stanford, CA 94309} \\ \vspace{0.1cm}\hrule\vspace{0.1cm}
Work supported in part by Department of Energy contract DE-AC02-76SF00515.
\end{center}

\section{Introduction}
\label{section-introduction}
The Standard Model (SM) of electroweak interactions describes \CP\ violation
(CPV) as a consequence of a complex phase in the
three-generation Cabibbo-Kobayashi-Maskawa (CKM) quark-mixing
matrix~\cite{ref:CKM}. Measurements of \CP\ asymmetries in
the proper-time distribution of neutral $B$ decays to
\CP\ eigenstates containing a charmonium and $K^{0}$ meson provide
a precise measurement of $\sintwob$~\cite{BCP}, where
$\beta$ is $\arg \left[\, -V_{\rm cd}^{}V_{\rm cb}^* / V_{\rm td}^{}V_{\rm tb}^*\, \right]$ and
the $V_{ij}$ are CKM matrix elements. The SM also predicts the amplitude of CPV in
$\b\rightarrow\c\cbar\d$ and $\bsqq$ $(\q=d,s)$ decays, defined as $\sin2\beta_{\rm eff},$ to be 
approximately $\sin2\beta .$ 
The $\bccd$ loop amplitudes have a different weak phase than 
the $\bccd$ tree amplitude and if there is a significant penguin amplitude 
in such $\bccd$ decays, then one will measure a value of $\sin2\beta_{\rm eff},$ 
that differs from $\sintwob$~\cite{grossman}. $\bsqq$ decays may also be especially 
sensitive to New Physics since they are dominated by one-loop transitions 
that can potentially accommodate large virtual particle masses and contributions from 
physics beyond the SM could invalidate this prediction~\cite{grossman}. However, many of these 
$\bsqq$ final states are affected by additional SM physics contributions that may obscure 
the measurement of $\beta_{\rm eff}$~\cite{grossman2}. Precise 
measurements of $\sin2\beta_{\rm eff}$ in many $\bccd$ and $\bsqq$ decays are therefore important
either to confirm the SM picture or to search for the possible presence of New Physics.
\section{Experimental Technique}
\label{section-experimental}
The \babar\ detector~\cite{ref:babar} is located at the SLAC PEP-II $e^+e^-$ asymmetric energy \B -factory.
Its program includes the study of CPV in the \B-meson system through the measurement of time-dependent 
$\CP$-asymmetries, $\Acp$. At the $\Upsilon(4S)$ resonance, $\Acp$ is extracted from the distribution of the 
difference of the proper decay times, $\t \equiv \tcp - \ttag$, where
$\tcp$ refers to the decay time of the signal \B\ meson ($\Bcp$) and $\ttag$ refers to the
decay time of the other \B\ meson in the event ($\Btag$). The decay products of $\Btag$ are
used to identify its flavour at its decay time. $\Acp$ is defined as:
\begin{equation}
\Acp(t) \equiv \frac{N(\Bzb(t)\to f_{CP}) - N(\Bz(t)\to f_{CP})} {N(\Bzb(t)\to f_{CP}) + N(\Bz(t)\to f_{CP})} = 
        \S \sin(\dm{t}) - \C \cos(\dm{t}),
\label{acpt}
\end{equation}
where $N(\Bzb(t)\to f_{CP})$ is the number of \Bzb\ that decay into the $CP$-eigenstate $f_{CP}$ after a time $t$.
$\Acp$ can also be expressed in terms of the difference between the \B\ mass eigenstates $\dm$, 
where the sinusoidal term describes the interference between mixing and decay and the cosine term is the 
direct \CP\ asymmetry. 
\section{Measurements of \sintwob\ from charmonium decays}
\label{section-charmonium}
The SM predicts that direct $\CP$ violation in $\bccs$ ($\bzero\rightarrow$ charmonium + $\kz$) decays 
is negligible. It follows that $\Acp(t)$ = $-$$\etaf\sintwob\sin$($\dm{t}$) where $\etaf$ 
is the eigenvalue corresponding to the $CP$-eigenstate $f_{CP}$. $\sintwob$ has been 
directly measured using $\bzero$ decays to the final states ${J}/\psi\ks$, $\psi\ks$, $\chiks$, 
$\etacks$, $\jpsikstar$($\kstardecay$) and ${J}/\psi\kl$~\cite{babar-jpsik}.
An extended unbinned maximum-likelihood (ML) fit to the data gives $\sintwob$ = 0.722 $\pm$ 0.040 $\pm$ 0.023 {\footnote{All 
results are quoted with the first error being statistical and the second being systematic.}}, which is 
in agreement with SM expectation. A four-fold ambiguity in $\beta$ that is obtained from this measurement 
is reduced to a two-fold ambiguity through the measurement of $\costwob$. 
Using 81.9 fb$^{-1}$ of integrated luminosity
$\costwob$ is measured as 2.72 $^{+0.50}_{-0.79}$ $\pm${0.27} using $\btojpsikstar$ decays~\cite{babar-jpsikstar}.
This determines the sign of $\costwob$ to be positive at 86$\%$ C.L. and is compatible with the sign 
of $\costwob$ inferred from SM fits of the unitarity triangle.
\section{Measurements of \sintwob\ from $\bccd$ decays}
\label{section-charm}
The decay $\btodstardstar$ is an admixture of $\CP$-odd and $\CP$-even components. By performing a
transversity analysis~\cite{babar-dstardstar}, the $\CP$-odd fraction is measured to be 
0.125 $\pm$ 0.044 $\pm$ 0.007. The time-dependent $\CP$ asymmetry parameters \S\ and \C\
are measured to be $-$0.75 $\pm$ 0.25 $\pm$ 0.03 and 0.06 $\pm$ 0.17 $\pm$ 0.03
respectively. A preliminary analysis of the decay $\btojpsipi$ also shows it to be consistent with the 
SM~\cite{babar-jpsipiz}. The signal yield, \S\ and \C\ are simultaneously 
extracted from a ML fit. 109 $\pm$ 12 events are measured with 
\C\ = $-$0.21 $\pm$ 0.26 $\pm$ 0.09 and \S\ = $-$0.68 $\pm$ 0.30 $\pm$ 0.04.
\section{Searches for New Physics}
\label{section-penguin}
Two $\bsqq$ $(\q=d,s)$ decays to \CP\ eigenstates that have been noted as having small theoretical
uncertainties in the measurement of $\beta_{\rm eff}$ are $\btophikz$ and 
$\btoksksks$~\cite{ref:gershon}. 
$\Bz$ decays to $\phi\ks$ and $\phi\kl$ are reconstructed and a ML
fit yields 114 $\pm$ 12 $\phi\ks$ and 98 $\pm$ 18 $\phi\kl$ $\Bz$ candidates.
 $\sintwob_{eff}$ is measured to be 0.50 $\pm$ 0.25 $^{+0.07}_{-0.04}$~\cite{babar-phiks}.
A ML fit of reconstructed $\Bz\rightarrow\ks\ks\ks$ candidates (where $\ks\rightarrow\pip\pim$), finds
\C\ = $-$0.34 $^{+0.28}_{-0.25}$ $\pm$ 0.05 and 
\S\ = $-$0.71 $^{+0.38}_{-0.32}$ $\pm$ 0.04~\cite{babar-ksksks-old}. A more recent
analysis, where one $\ks$ is reconstructed in the $\ks\rightarrow\piz\piz$ mode,
was combined with~\cite{babar-ksksks-old} to give the preliminary results: 
\C\ = $-$0.10 $\pm$ 0.25 $\pm$ 0.05 and \S\ = $-$0.63 $^{+0.32}_{-0.28}$ $\pm$ 0.04~\cite{babar-ksksks}.
The experimental challenge in~\cite{babar-ksksks} came from the absence of charged tracks originating 
from the $\bzero$ decay vertex~\cite{babar-kspiz}. 
\par The decay $\btoetaprimekz$ is also interesting, since additional contributions estimated using SU(3) and 
QCD factorisation are expected to be small~\cite{etaprimetheory}. A ML fit to reconstructed 
$\btoetaprimekl$ and $\btoetaprimeks$ candidates yields the preliminary result of 1245 $\pm$ 67 candidates, 
\S\ = 0.36 $\pm$ 0.13 $\pm$ 0.03 and \C\ = $-$0.16 $\pm$ 0.09 $\pm$ 0.02. 
The value of \S\ = $\sin2\beta_{\rm eff}$ differs from the \babar\ value of $\sintwob$ as measured 
in charmonium + $\kz$ decays by 2.8 standard deviations~\cite{babar-etaprimekz}. 
Other $\bsqq$ decays have been studied at $\babar$. These include $\btofzkz$, $\btopizkz$, $\btopizpizkz$, 
$\btoomegakz$ and $\btokpkmkz$~\cite{hfag,b2speng}. Small deviations from SM expectations are seen.
\section{Conclusion}
\label{section-conclusion}
$\sintwob$ has been measured to 5$\%$ accuracy using $\bzero\rightarrow$ charmonium + $\kz$ decays
and is consistent with SM expectations. No deviation from the SM has been observed in $\bccd$
decays. Future updates of the $\bsqq$ analyses on larger datasets will help to understand if the present 
pattern in the deviation of $\b\rightarrow\s$ penguins from SM predictions is a statistical effect 
or a sign of New Physics.

\bibliographystyle{aipproc}   

\end{document}